\documentclass[twoside]{dis08}
\usepackage[dvips]{graphicx,epsfig,color}
\usepackage{wrapfig,rotating}
\usepackage{amssymb,amsmath,array}

\begin{document}
\title{
Summary from the working group on ``Diffraction and Vector Mesons'' at DIS\,2008
}

\author{Marcella Capua$^1,$ Monika Grothe$^2,$ Dmitry Ivanov$^3,$ and
Mikhail Kapishin$^4$
%
%
\vspace{.3cm}\\
%
1- Calabria University and INFN - Dept of Physics \\
Arcavacata di Rende (CS) - Italy - \texttt{capua@cs.infn.it}\\
\vspace{.1cm}\\
2- University of Wisconsin, Madison  - Department of Physics \\
1150 University Ave., Madison, WI 53706  - USA - \texttt{monika.grothe@cern.ch}\\
\vspace{.1cm}\\
3- Sobolev Institute of Mathematics and Novosibirsk State University
\\ 630090 Novosibirsk - Russia -
\texttt{d-ivanov@math.nsc.ru}\\
\vspace{.1cm}\\
4- Joint Institute for Nuclear Research - Laboratory of High Energy Physics \\
Joliot Curie 6, 141980 Dubna, Russia - \texttt{kapishin@mail.desy.de}
}

\newcommand{\spom} {\mbox{$\scriptstyle \mathrm{I}\! \mathrm{P}$}}
\newcommand{\xpom} {\mbox{$x_{\spom}$}}
\maketitle

\begin{abstract}
The contributions at the DIS2008 workshop in the working group on Diffraction and Vector
Mesons are summarised.
\end{abstract}

\section{Introduction}

In the 1960s, one of the mysteries of strong interactions was the enormous proliferation of
strong interacting particles or hadrons, which in addition, exhibit a striking property:
the more massive the particles, the higher their spin with a linear correlations between
the square of the particle mass and its spin. In Regge Theory~\cite{regge}
this correlation is
described in the complex angular momentum plane by linear Regge trajectories.
In this phenomenological approach diffraction corresponds to the exchange of the
Pomeron trajectory, which has the quantum numbers of the vacuum.
Pomeron trajectory exchanges dominate at high energies and its parameters have been
derived from fits to the data of soft hadron-hadron interactions~\cite{dl}.

The study of the hadronic structure performed at large centre-of-mass energies in $p\bar p$ collisions at the Tevatron and in deep inelastic $ep$  scattering (DIS) at HERA offer unique way to understand diffraction in terms of Quantum Chromo-Dynamics (QCD)(see
e.g.~\cite{arneodo} and
references therein).
Within the QCD framework diffractive events can be interpreted as processes in which a colour
singlet combination of partons is exchanged.

The structure of the colour singlet can be studied using a QCD approach based on the hard scattering
collinear factorization theorem \cite{fact}. It states that the diffractive
cross section is a product of diffractive proton parton distribution functions
(DPDFs) and partonic hard scattering cross sections. DPDFs
are universal for diffractive $ep$ DIS processes (inclusive, dijet and
charm production) and obey the DGLAP evolution equations.
Their extraction from inclusive diffractive processes and application to the analysis of diffractive dijet and charm production provides a test of QCD factorisation. Moreover, dijet and charm production data, directly sensitive to the gluon density in the colour singlet, can help in constraining the DPDFs when included in global fits.

\section{Diffraction in $ep$ interactions}

The H1 and ZEUS experiments at HERA study processes of
electron-proton DIS,  $ep\rightarrow eX$, where $X$ is the hadronic final state.
A considerable part of the DIS cross section ($\sim10\%$) comes from diffractive processes which are characterized by the presence of a leading proton in the final state carrying most of the proton beam energy and by a large rapidity gap (LRG) in the forward (proton) direction.

\begin{wrapfigure}{r}{0.25\columnwidth}
\vspace{-0.4cm}
\centerline{\includegraphics[width=0.20\columnwidth]{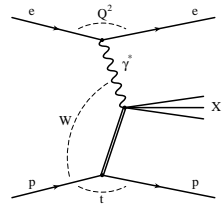}}
\caption{Diagram of the diffractive process $ep\rightarrow eXp$.}
\label{Fig:fig1}
\end{wrapfigure}

Figure~\ref{Fig:fig1}  shows a schematic diagram of the diffractive process $ep\rightarrow eXp$
together with the relevant kinematic variables: the photon virtuality, $Q^2$, the photon-proton centre-of-mass energy, $W$, and the square of the four-momentum transfer at the proton vertex, $t$.
Relevant kinematic variables in diffraction are the fraction of the proton longitudinal momentum carried by the exchanged colour singlet object,  $\xpom$,  and the fraction of the exchanged momentum carried by the quark coupling to the virtual photon, $\beta$.

In analogy with the structure function $F_2$ for inclusive DIS the cross section of diffractive $ep$ scattering is parameterized in terms of the diffractive structure function $F_2^{D(4)}(\beta,Q^2,\xpom,t)$.

\subsection{Inclusive and semi-inclusive diffraction in $ep$ interactions}

Diffractive processes at HERA are selected in several ways: by requiring the presence of a large rapidity gap (LRG method~\cite{LRG}), by observing a leading proton in the final state by means of a dedicated leading proton spectrometer (LP method~\cite{ptagging}) or by studying the hadronic mass spectrum observed in the central detector ($M_X$ method~\cite{mx,mx1}) to determine the diffractive contribution statistically.

The different methods treat in different
ways the contribution of non-diffractive processes, like exchanges of the sub-leading Regge trajectories, and the background from proton dissociation processes.

The ZEUS experiment presented~\cite{ruspa} the diffractive inclusive DIS results obtained with all the three experimental methods~\cite{mx2,marta}, and their comparison to H1 recent results obtained with the LRG and LP methods~\cite{h1lrg,h1fps}.
The results obtained with the three methods were compared in bins of $M_X$, $Q^2$ and $\xpom$ in terms of the diffractive reduced cross section, $\sigma_r^{D(4)}(\beta ,Q^2, \xpom ,t)$. The latter coincides with the diffractive structure function, $F_2^{D(4)}$, if the ratio of the cross sections for longitudinally and transversely polarised virtual photons is neglected. The reduced cross section  $\sigma_r^{D(3)}(\beta ,Q^2, \xpom )$ is integrated over $t$ for the LRG and $M_X$ methods which do not allow a direct measurement of $t$, possible only with the LP method.

The LRG and LP results of ZEUS cover kinematic range: $Q^2>2$~GeV$^2$, $M_X>2$~GeV, $40<W<240$~GeV and proton fractional momentum
losses $0.0002<\xpom <0.02$ (LRG) or $0.0002<\xpom <0.1$ (LP).
With the LRG and $M_X$ methods high $M_X$ values are not accessible. Events in which proton dissociates into a low mass state give
contribution to the measured cross section.
A detailed study of the events of proton dissociation allowed ZEUS to estimate their fraction
in the LRG inclusive sample as $25 \pm 1 {\rm (stat.)} \pm 3 {\rm (syst.)}\%$.
In the LP method the proton dissociation background is negligible for $\xpom < 0.02$.

\begin{figure}[htb]

\vspace{-0.2cm}
\includegraphics[height=.35\textheight]{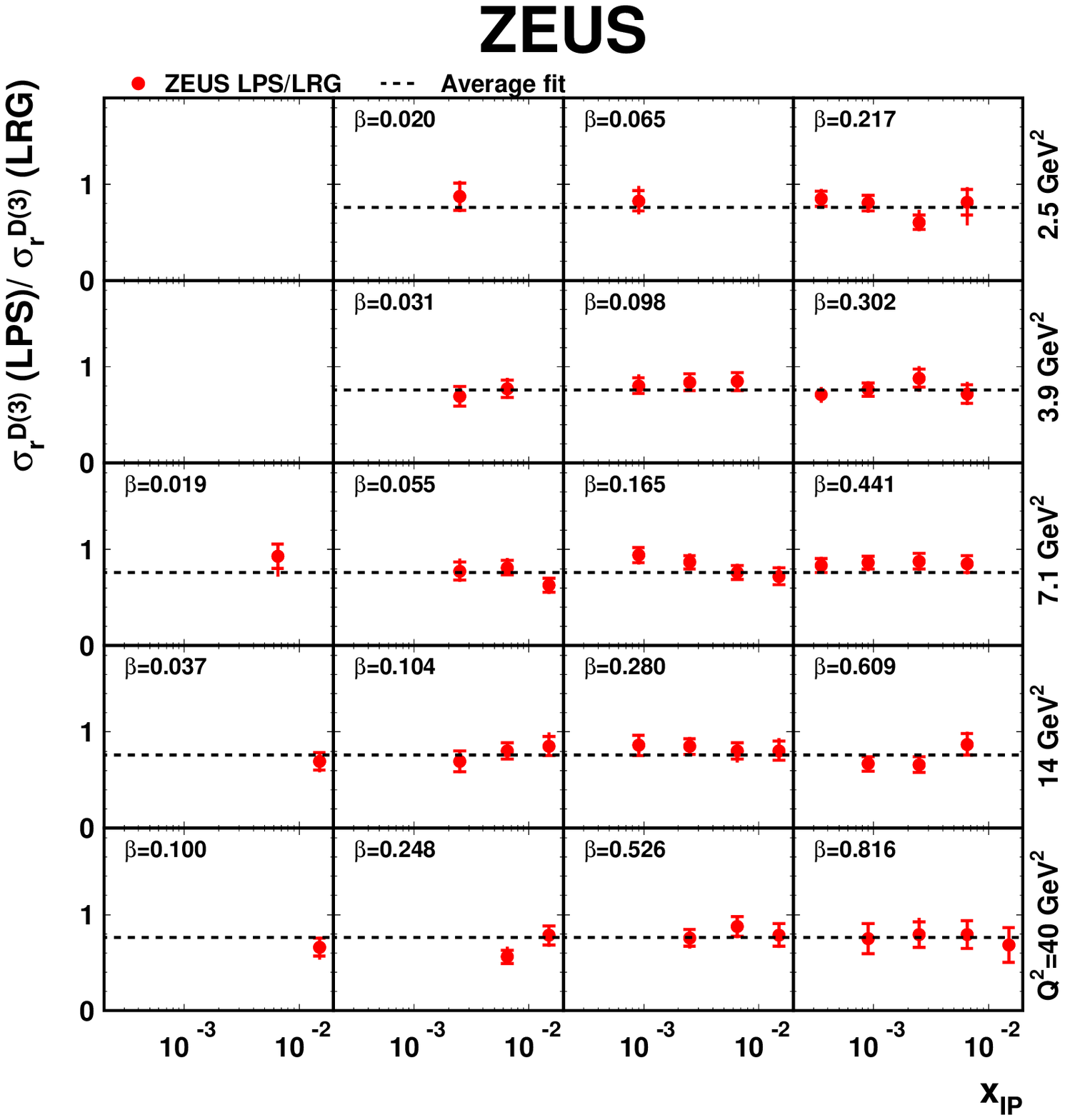}
\includegraphics[height=.35\textheight]{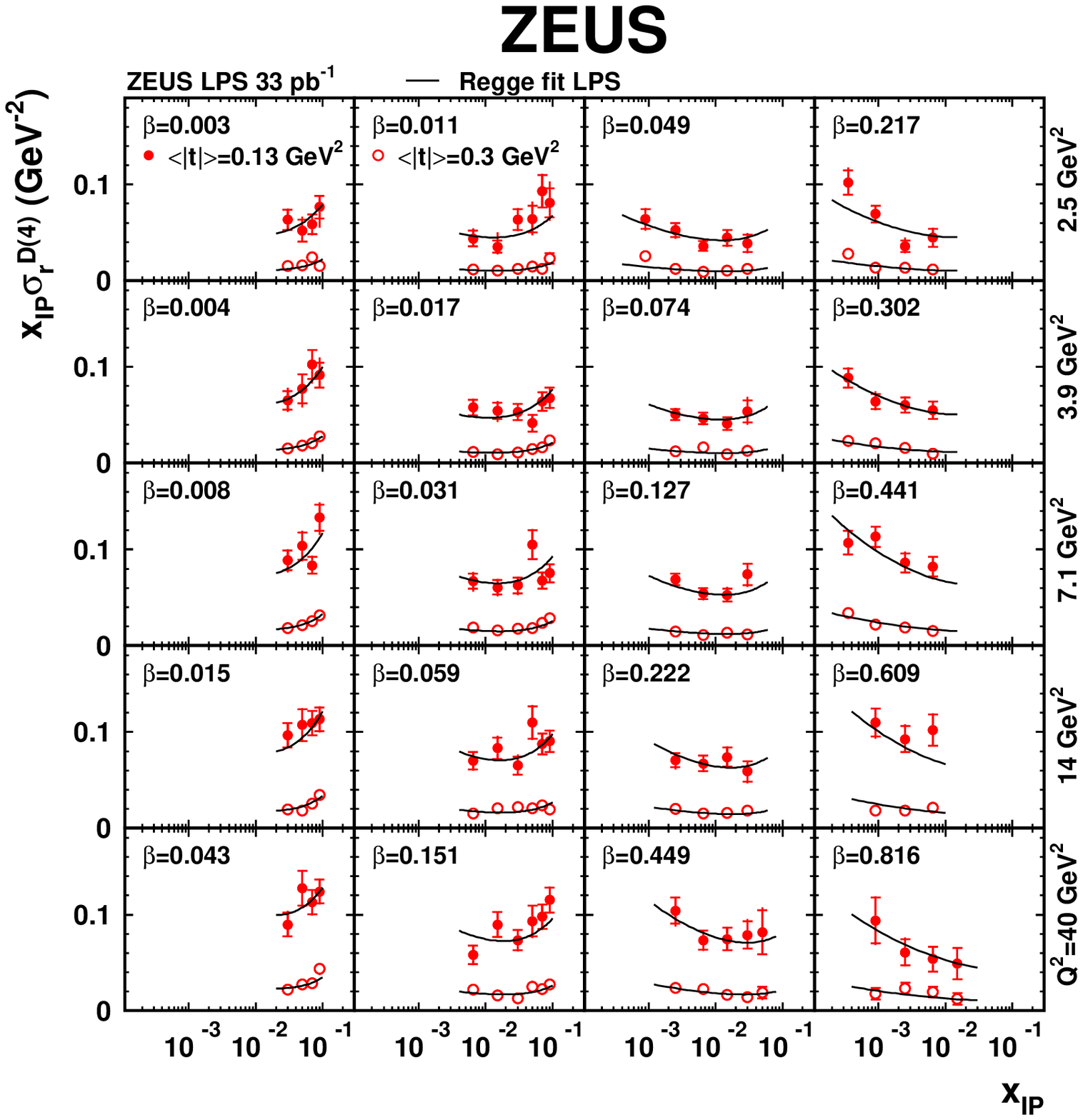}

\vspace{-0.3cm}
\caption{Ratio of the reduced diffractive cross sections
obtained by ZEUS with the LP and the LRG methods as
a function of $\xpom$ for different values of $Q^2$ and $\beta$.
(left). Reduced diffractive cross section obtained with the LP method
measured in two $t$ bins as
a function of $\xpom$ for different values of $Q^2$ and $\beta$ (right).
The data are compared with the results of a Regge fit. }
\label{Fig:fig2}
\vspace{-0.3cm}
\end{figure}

\begin{figure}[htb]

\vspace{-0.2cm}
\includegraphics[height=0.35\textheight]{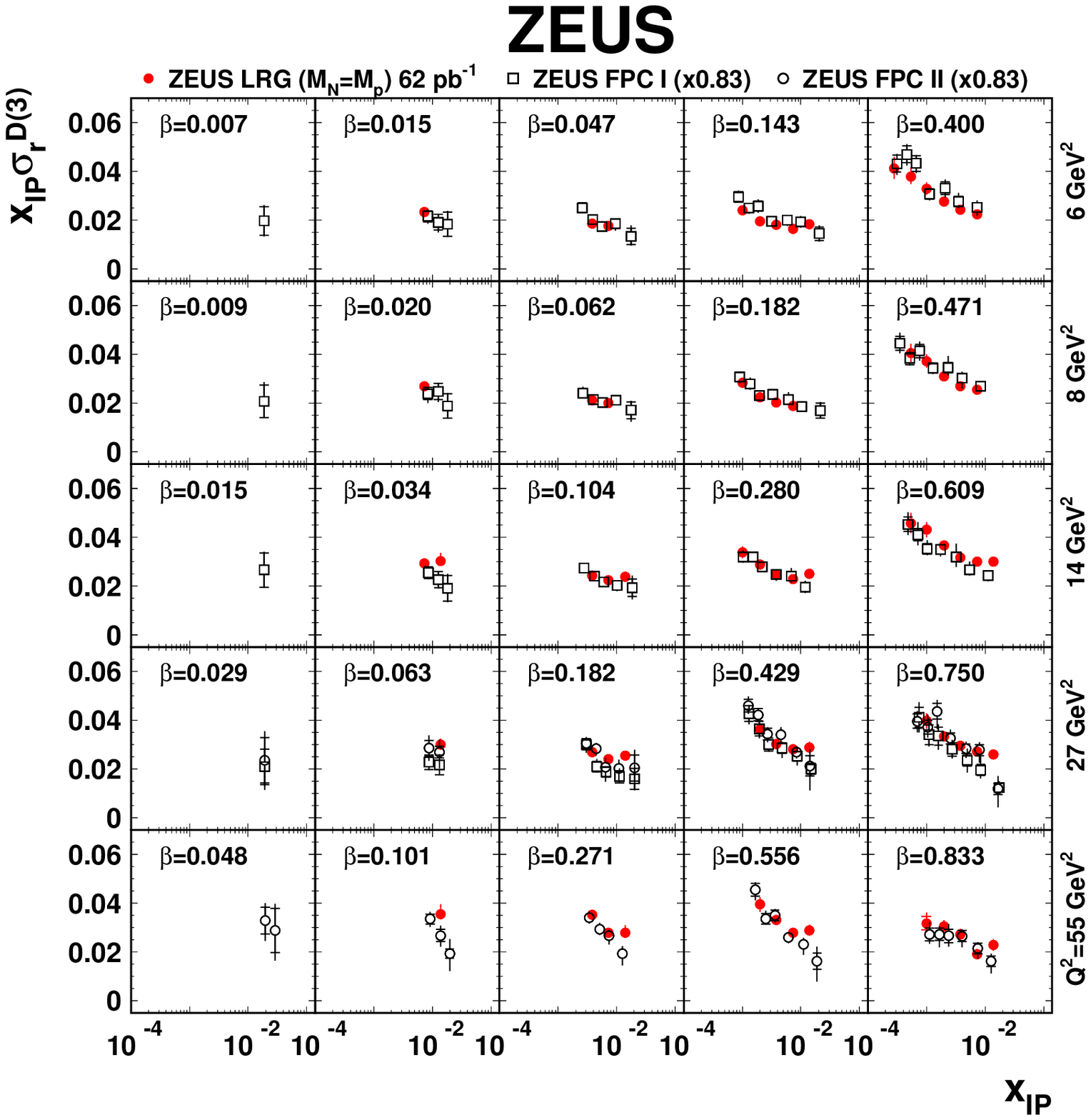}
\includegraphics[height=0.35\textheight]{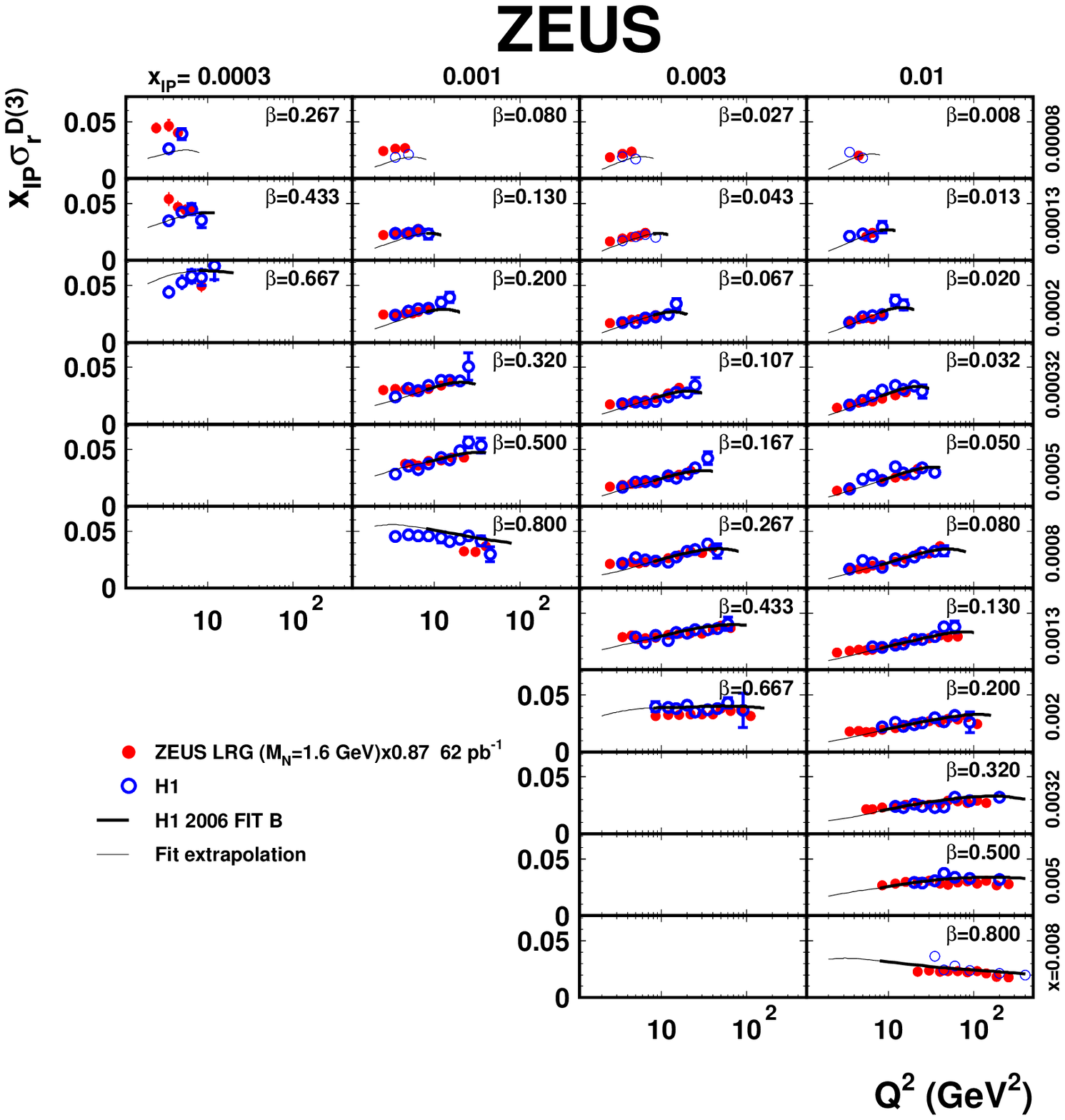}

\vspace{-0.3cm}
\caption{(left) Reduced diffractive cross section obtained by ZEUS with the LRG and $M_X$ methods as a function of $\xpom$ for different values of $Q^2$ and $\beta$.
(right) Reduced diffractive cross section obtained with the LRG method  as a function of $Q^2$ for different values of $\xpom$ and $\beta$ compared with the H1 results and with the H1 DPDF 2006 Fit B predictions, the ZEUS results are normalised to the H1 results.}
\label{Fig:fig3}

\vspace{-0.5cm}
\end{figure}

The ratio of the reduced diffractive cross sections,
$\sigma_r^{D(3)}$, obtained with the LP and LRG methods
is found to be independent of $Q^2$, $\xpom$ and $\beta$ (Fig.~\ref{Fig:fig2},left). A similar comparison has been performed by H1,
also showing a good agreement~\cite{newmann}.

The ZEUS experiment has also shown the first measurement, obtained with the LP method
in two $t$ bins, of the reduced cross section $\sigma_r^{D(4)}$ as a function
of $\xpom$ (Fig.~\ref{Fig:fig2},right).
The results show the same $\xpom$ dependence for each $Q^2$ and $\beta$ bin.
A Regge fit indicates that the shape of the $\xpom$ dependence
is the same in two $t$ bins.

A satisfactory agreement between the data obtained with the LP method in both the
ZEUS and H1 experiments has been observed.
The ZEUS and H1 results are consistent with the hypothesis of proton vertex factorization~\cite{ingsc}.
Figure~\ref{Fig:fig3} (left) shows a reasonable agreement of the ZEUS results obtained with the $M_X$ and LRG methods.
The different $\xpom$ dependence in some bins at low $Q^2$ may be ascribed to the contribution
of the sub-leading Reggeon and pion trajectories which is suppressed in the $M_X$ method and not suppressed in the LRG method.
A comparison of the results obtained with the LRG method by ZEUS and the analogous results from H1, fig.~\ref{Fig:fig3} (right), shows a reasonable agreement.

The H1 Collaboration presented~\cite{laycock} the results on the diffractive PDFs~\cite{h1lrg}, extracted from the H1 LRG inclusive measurements for $Q^2>8.5$~GeV$^2$, $M_X\geq 2$~GeV and $\beta \leq 0.8$.
The quark singlet and gluon distributions are extracted from NLO QCD fits to the data, the H1 2006 DPDF Fit A and B. These two fits differ in the parameterisation chosen for the gluon density at the starting scale for the QCD evolution. The resulting gluon distribution carries an integrated fraction of $\sim70\%$ of the momentum of the diffractive exchange.

\vspace*{2cm}
\begin{figure}[htb]

\vspace{-2.2cm}
\includegraphics[height=0.04\textheight]{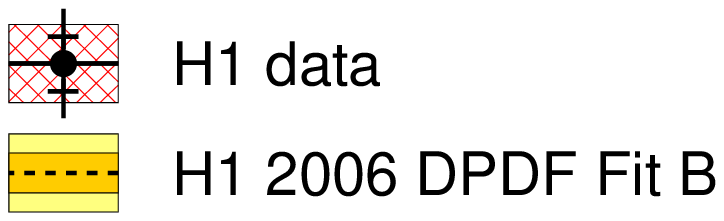}

\includegraphics[height=0.25\textheight]{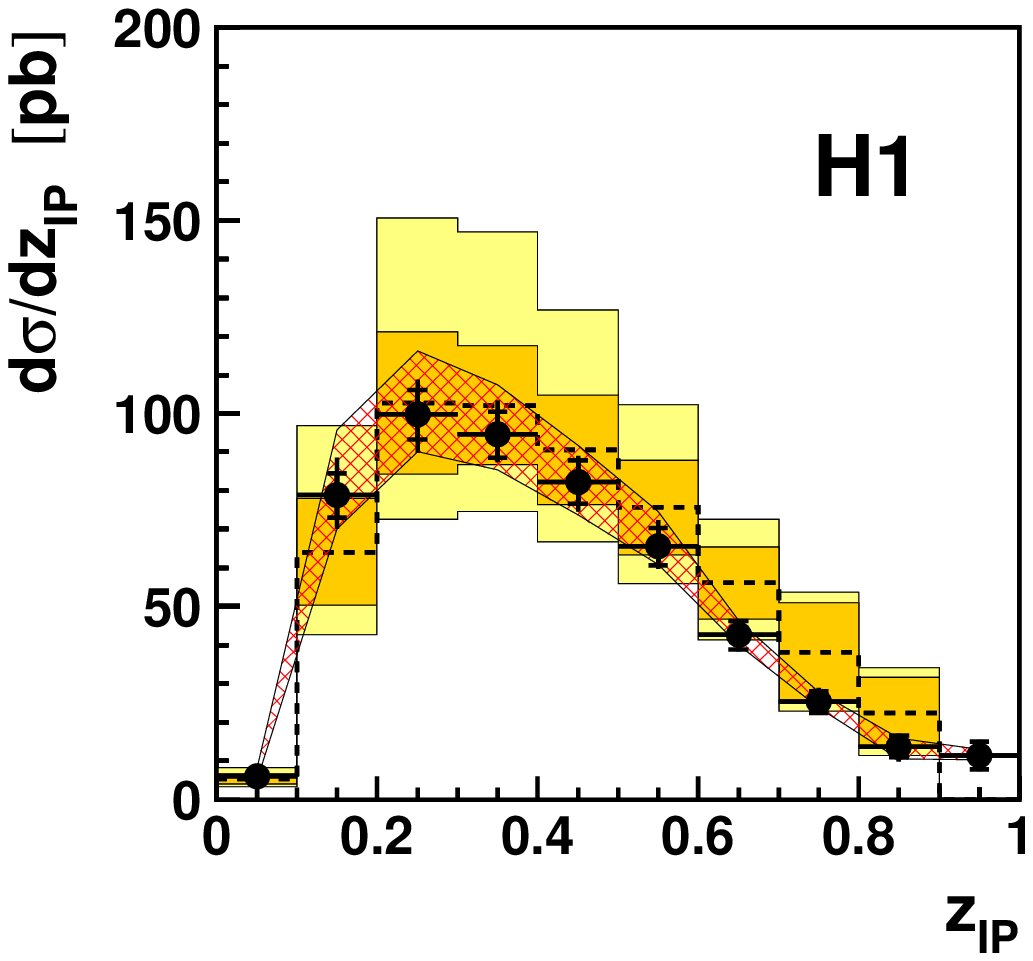}

\vspace{-4cm}
\hspace{5.4cm}
\includegraphics[height=0.08\textheight]{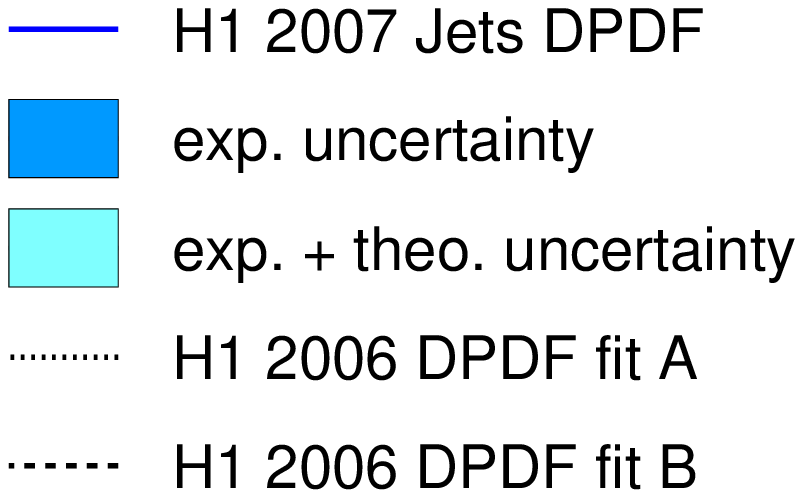}

\vspace{-2.5cm}
\hspace{8.2cm}
\includegraphics[height=0.24\textheight]{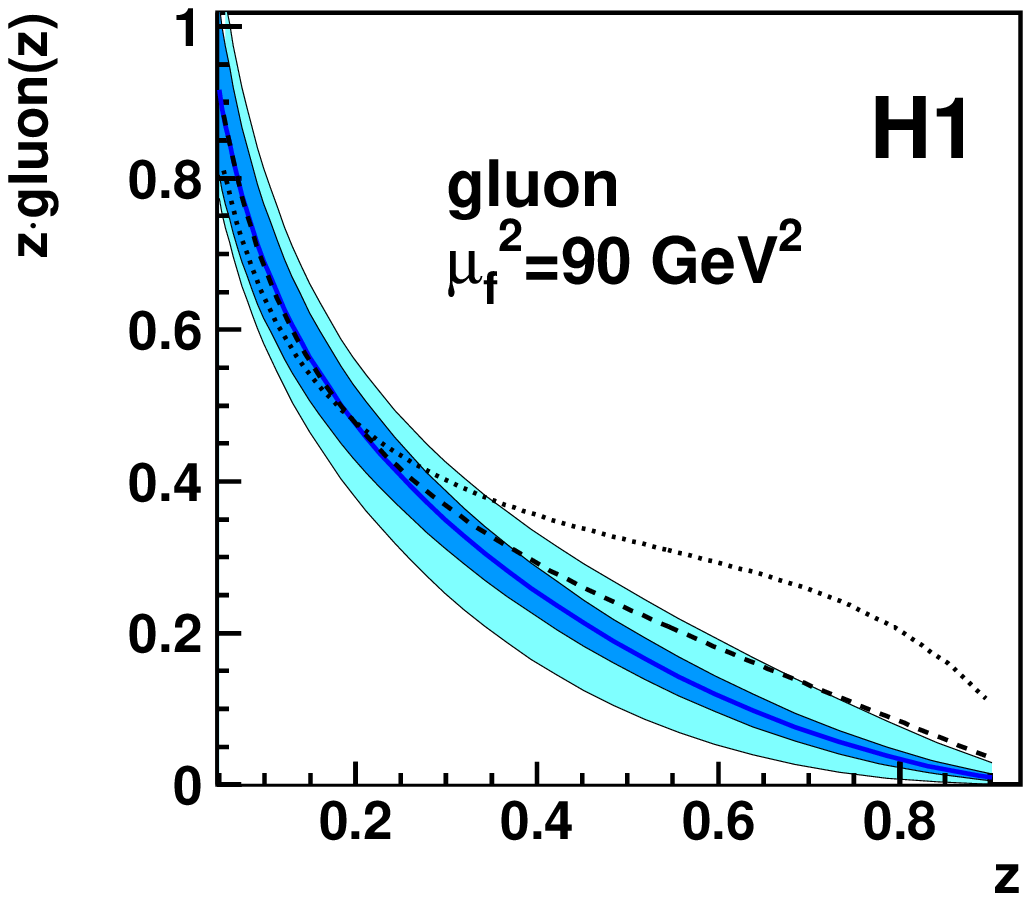}

\vspace{-0.3cm}
\caption{(left) Differential cross section of diffractive dijet production in DIS vs $z_{IP}$ compared with H1 2006 DPDF fit B. (right) Diffractive gluon density for a value of the hard scale of $90~{\rm GeV}^2$. The solid line indicates the combined fit of the H1 inclusive and dijet DIS data, the dashed lines shows the results of the H1 2006 DPDF fit A and B to inclusive DIS data.}
\label{Fig:fig5}
\vspace{-0.2cm}
\end{figure}

The H1 DPDF results has been used to predict the diffractive dijet cross sections in DIS~\cite{h1pdf}.
This process is dominated by boson-gluon fusion, where a hard
collision of a virtual photon and a gluon produces a high-$p_T$ $q\bar q$ pair.
Therefore, the diffractive dijet data are directly sensitive to the gluon
content of the diffractive exchange.

Figure~\ref{Fig:fig5}(left) shows the differential cross section for diffractive dijet production in DIS as a function of $z_{\spom}$, the fraction of the momentum of the diffractive exchange carried by the parton entering the hard scattering, compared to a prediction based on the H1 2006 DPDF Fit B. The $z_{\spom}$ distribution is the most sensitive to the gluon DPDFs.
The H1 2006 DPDF Fit B describes the diffractive dijet DIS cross section, whereas the Fit A overestimates the dijet data.  The dijet data were used by H1 as an additional constrain for both quark singlet and gluon densities in the NLO QCD procedure.
The new set of diffractive PDFs, the H1 2007 Jets DPDF, provides the most precise diffractive quark and gluon distributions in the range $0.05<z_{IP}<0.9$. The simultaneous fit of both the inclusive and dijet data provides big improvement in the precision of the gluon density at high fractional momentum compared to the previous extractions, fig.~\ref{Fig:fig5} (right).

In dijet photo-production (PhP) the hard scale is defined by $E_T$ of jets because
$Q^2 \sim 0$. QCD collinear factorization is expected to be valid in direct processes with pointlike photon, but broken in processes with resolved photons, where secondary
interactions between photon and proton remnants fill the rapidity gap.
These two processes can be separated using the variable $x_{\gamma}$, which corresponds to the longitudinal momentum fraction
of the photon entering the hard sub-process. Resolved photon processes correspond to $x_{\gamma}<1$, whereas direct photon
processes to $x_{\gamma} \simeq 1$.

\begin{wrapfigure}{r}{0.38\columnwidth}

\vspace{-0.5cm}
\centerline{\includegraphics[width=0.32\columnwidth]{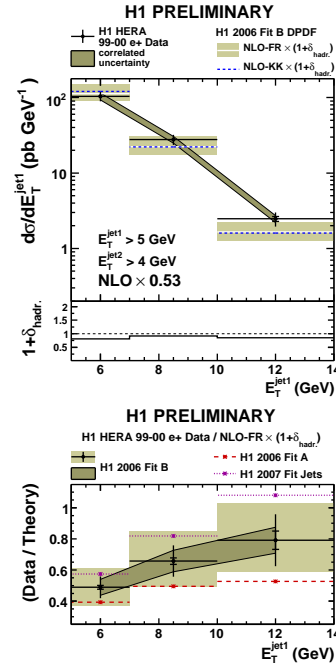}}

\vspace{-0.2cm}
\caption{Differential cross section of the diffractive dijet photo-production as a function of $E_T$ of the hardest jet. The data are compared to the NLO QCD predictions based on the H1 DPDF fits.}
\label{Fig:fig5c}
\vspace{-0.6cm}
\end{wrapfigure}

Neither H1~\cite{cerny}, nor ZEUS~\cite{slominski} observe a significant difference in the description of the data in the resolved (low $x_{\gamma}$) and direct (high $x_{\gamma}$) regions by  NLO predictions based on QCD collinear factorization and
DPDFs. The H1 Collaboration has also presented measurements in two ranges of $E_T$ of jets and found that the suppression of the dijet data relative to NLO
predictions is decreasing with increasing of $E_T$ of jets (Fig.~\ref{Fig:fig5c}). The ZEUS results also confirm this tendency. The dijet photoproduction cross sections from H1 and ZEUS are found to be consistent within uncertainties when measured in the same $E_T$ range, although the ratio of the H1 cross section to NLO predictions is somewhat lower than the ZEUS value.

\subsection{Exclusive diffraction in $ep$ interaction }

The transition from purely soft to hard pQCD processes is studied at HERA in the vector meson (VM) production. In particular, it is expected that the VM production cross section increases with increasing photon-proton centre-of-mass energy as $\sim W^{\delta}$, with the power factor $\delta$ growing with $Q^2$. Moreover, the effective size of the virtual photon decreases with $Q^2$, leading to a flatter distribution in $t$. Both the features have been observed at HERA and presented at the conference.

\begin{figure}[htb]

\vspace{0.5cm}
\hspace{1.0cm}
\includegraphics[height=.25\textheight]{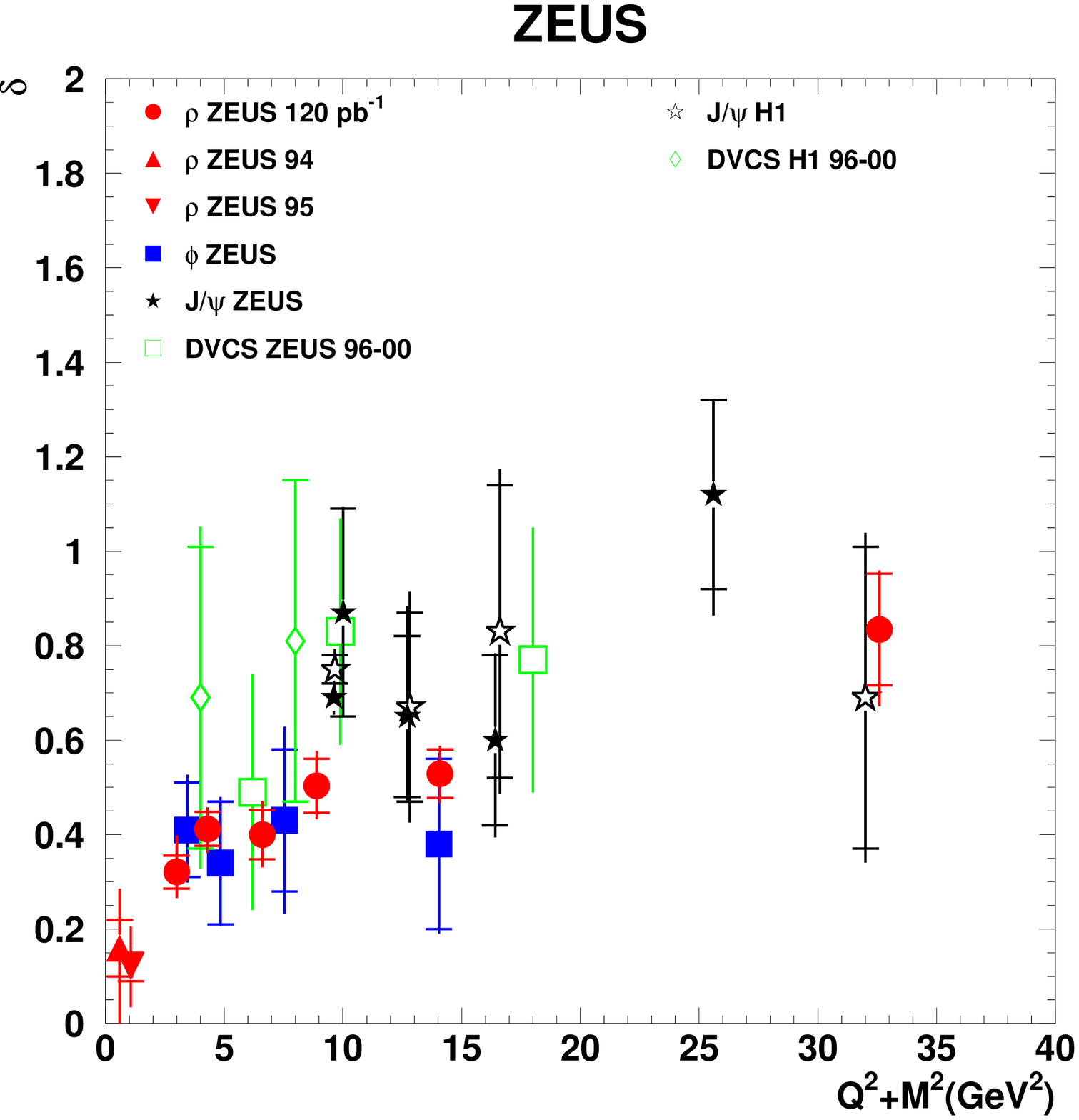}

\vspace{-5.1cm}
\hspace{6.5cm}
\includegraphics[height=.28\textheight]{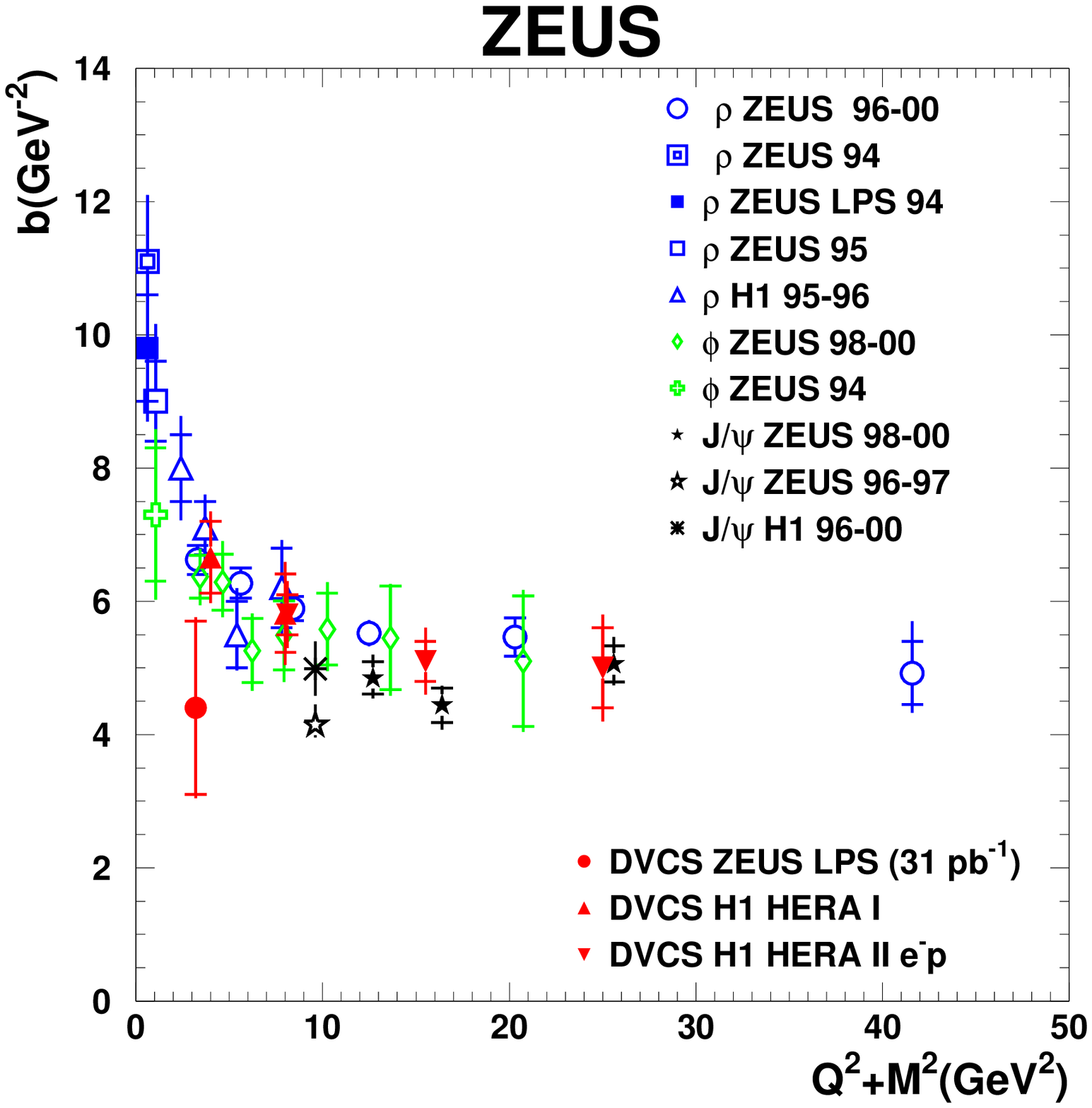}
\vspace{-0.3cm}
\caption{A compilation of the values of $\delta$ (left) and $b$ (right) for exclusive VM electroproduction as a function of $Q^2+M^2$ including the DVCS results.}
\label{Fig:fig6}
\vspace{-0.3cm}
\end{figure}

In particular, H1 presented new results on the $\rho$ and $\phi$ electro-production~\cite{janssen} and DVCS~\cite{favart,h1dvcs}). ZEUS presented recently published results on the $\rho$-meson electro-production~\cite{ukleja} and new results on $\Upsilon$ in photoproduction and DVCS~\cite{malka,marcella}.

\begin{wrapfigure}{r}{0.40\columnwidth}

\vspace{-0.5cm}
\centerline{\includegraphics[width=0.35\columnwidth]{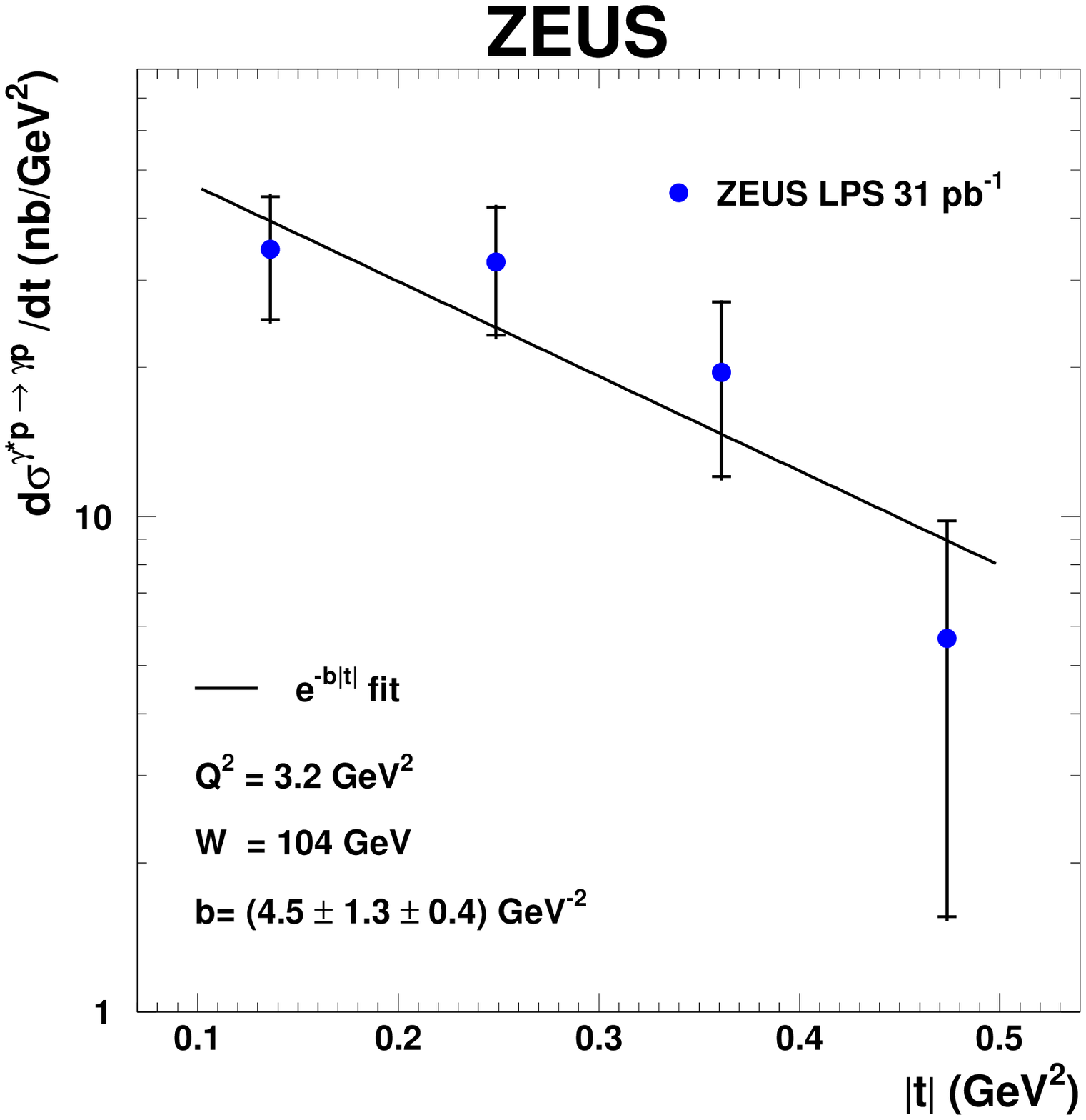}}
\centerline{\includegraphics[width=0.35\columnwidth]{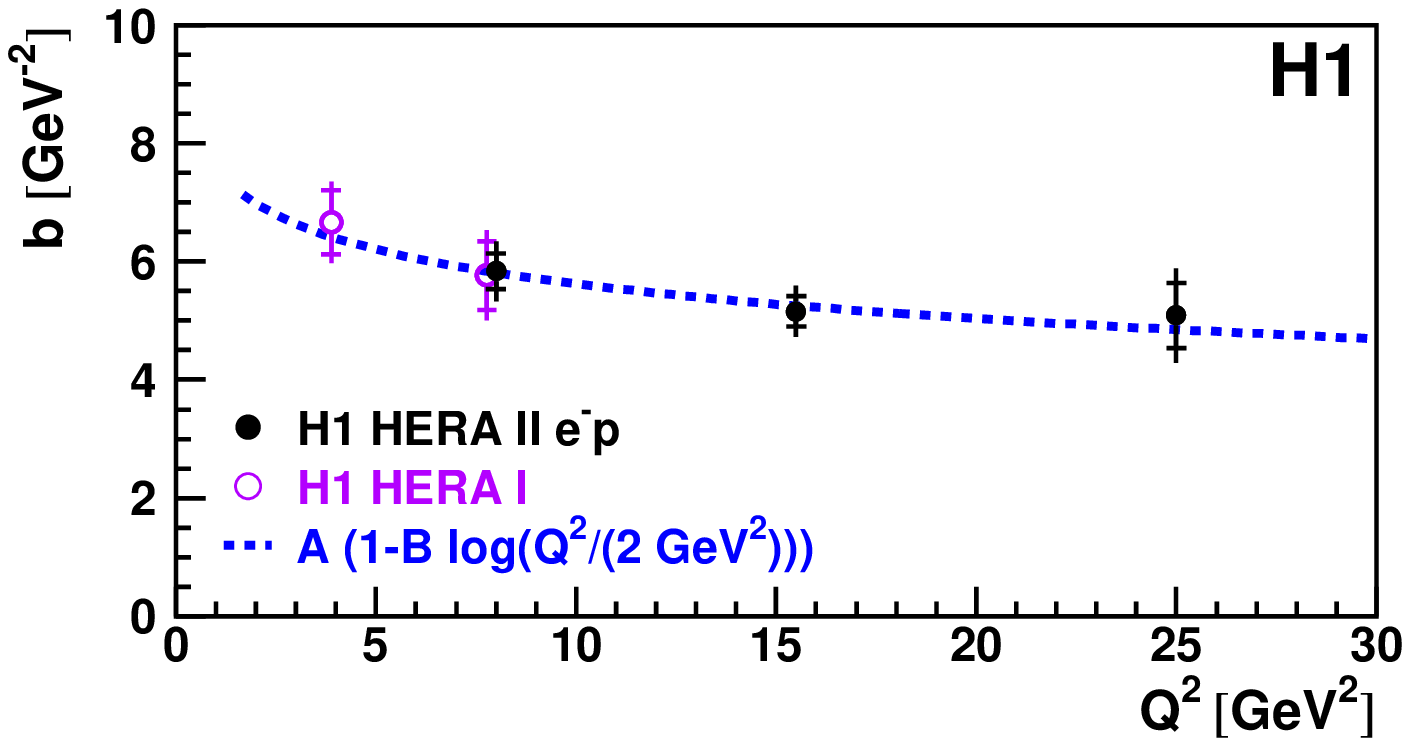}}
\caption{Differential cross section of DVCS measured by ZEUS a function of $t$. The result of the fit with an exponential function is shown (upper). Parameter $b$ of the DVCS $t$-distribution measured by H1 as a function of $Q^2$ (lower).}
\vspace{-0.4cm}
\label{Fig:fig7}
\end{wrapfigure}

The H1 and ZEUS results are summarized in Fig.~\ref{Fig:fig6}. The left figure shows a compilation of the $\delta$ values as a function of the scale $Q^2+M^2$ obtained for various VM and DVCS. It is found that $\delta$ increases with the scale for all VMs, as expected in pQCD. The right figure shows a compilation of the $b$ values obtained from an exponential fit to the differential cross section as a function of $t$. The values are presented as a function of the scale $Q^2+M^2$ for various VMs and DVCS. The figure shows the transition from soft to hard regime with $b$ decreasing with increasing scale up to the asymptotic value of 5~GeV$^{-2}$, as expected in pQCD.

The $t$-dependence of the DVCS cross section has been studied by ZEUS using a direct measurement of the variable $t$ with a LP spectrometer. The measurement was performed in the kinematic region, $Q^2>1.5$GeV$^2$ and $40<W<170$GeV. The result is presentsd in the upper part of Fig.~\ref{Fig:fig7}.
In the DVCS analysis of H1 the variable $t$ was computed as a vector sum of the transverse momenta of the final state photon and the scattered lepton. H1 studied the differential cross section as a function of $t$ for different values of $Q^2$ and $W$. The lower part of Fig.~\ref{Fig:fig7} shows the $b$ values obtained by H1 for different $Q^2$ values. A soft $Q^2$ dependence and no $W$-dependence of the parameter $b$ was observed.

H1 also measured the ratio of the proton dissociative to elastic cross sections for $\rho$ and $\phi$ electroproduction~\cite{janssen}. The ratio is found to be independent of $Q^2$. The result is consistent with the hypothesis of proton vertex factorisation. The $Q^2$ and $t$ dependences of spin density matrix elements have been studied and the ratios of polarised amplitudes in $\rho$ and $\phi$ electroproduction have been extracted confirming SCHC violation.

\section{Results from the Tevatron and prospects at the LHC}


Describing hard diffractive hadron-hadron
collisions is more challenging than describing hard diffractive $ep$ collisions since factorisation is broken
by
rescattering between spectator partons. These rescattering effects, often
quantified in terms of the so-called ``rapidity gap survival
probability''~\cite{Bj}, $\langle | S^2| \rangle$, are of interest in their own right because of
their relation with multiple parton scattering. To first approximation, the cross section for
diffractive hadron-hadron collisions is directly proportional to $\langle | S^2| \rangle$,
independent of kinematics.

At the Tevatron, single diffraction cross sections for vector boson, di-jet and heavy quark production
are observed that are lower by a factor $\mathcal{O}(10)$ relative to the
theoretical predictions based on the diffractive PDFs from HERA. This translates into a
ratio of diffractive to non-diffractive production at the Tevatron of $\sim 1 \%$.
Theoretical expectations at the LHC are at the level of a fraction of a per cent~\cite{CF}.
There are, however, significant uncertainties in the predictions, notably due to the
uncertainty of $\langle | S^2| \rangle$. While there is some consensus that
$\langle | S^2| \rangle \simeq 0.05$ \cite{Ssquared}, values between 0.004 and 0.23 have been
proposed \cite{miller}.

Interest in diffraction at the LHC has been greatly boosted recently by theoretical
predictions \cite{CEP-KMR}
that identified central exclusive production (CEP) as a potential discovery channel for the
Higgs boson. Experimental evidence from the Tevatron on CEP
of di-jets, di-photons and di-leptons support the validity of the undelying theoretical
model.
In the last years, both ATLAS and CMS have set up forward physics programs. In addition,
FP420, a joint R\&D program of ATLAS, CMS and the LHC machine group has investigated the
feasibility of an upgrade of the forward detector instrumentation to make possible the direct
observation of the scattered protons in CEP of a Higgs boson.

\subsection{Diffraction with a hard scale}

The CDF experiment reported preliminary results \cite{goulianos} from their run-II data for the
process $\bar{p} p \rightarrow p W X$ and $\bar{p} p \rightarrow p Z X$.
These processes are sensitive to the quark component of the diffractive PDFs.
The run-II analysis employs the MiniPlug calorimeters ($ 3.5 < | \eta| < 5.5$) for $E_T$ and angular measurements, Beam Shower Counters ($ 5.5 < | \eta | < 7.5$) to identify rapidity gaps
and the CDF Roman Pot Spectrometer to
detect leading $\bar{p}$ with $0.03 < \xi < 0.09$, where $\xi$ is the fractional momentum
loss of the $\bar{p}$.
A novel feature of the analysis is the determination of the full kinematics of the
$W\rightarrow e\nu/\mu\nu$ decay by obtaining the neutrino $E_T^\nu$ from the
missing $E_T$, as usual, and $\eta_\nu$ from the difference of the $\xi$ measured by the
Roman Pot detectors and the central calorimeter.
The results for the ratio of single diffractive to non-diffractive production
obtained from a data sample of $0.6 ~ \rm{fb}^{-1}$ are:

$R_W(0.03<\xi<0.10,\,|t|<0.1)=[0.97\pm 0.05\;\mbox{(stat)}\pm 0.11\;\mbox{(syst)]}\%$

$R_Z(0.03<\xi<0.10,\,|t|<0.1)=[0.85\pm 0.20\;\mbox{(stat)}\pm 0.11\;\mbox{(syst)]}\%$ \newline
These results are in good agreement with the run-I results from D0 and CDF
after correcting for differences in acceptance to the rapidity-gap-based selection used then.

The CMS experiment studied the feasibility of observing single diffractive W
production \cite{vilela}
with $W \rightarrow \mu \nu$ at
LHC energies and for an effective integrated luminosity for single interactions of
$100 ~ \rm{pb}^{-1}$.
The selection is rapidity-gap based and requires the absence of
activity in the forward calorimeters of CMS, i.e. in HF ($3 < | \eta | < 5$) and CASTOR
($-6.6  < \eta  < -5.2$). At LHC start-up the latter will be available only on one side of the
interaction point. Assuming a rapidity gap survival probability of 0.05,
$\mathcal{O}(100)$ signal events are expected with a
signal-to-background ratio as high as 20 if the CASTOR calorimeter is
available.

\subsection{Central exclusive production processes at the Tevatron}

\begin{figure}[ht]
 \includegraphics[width=4cm]{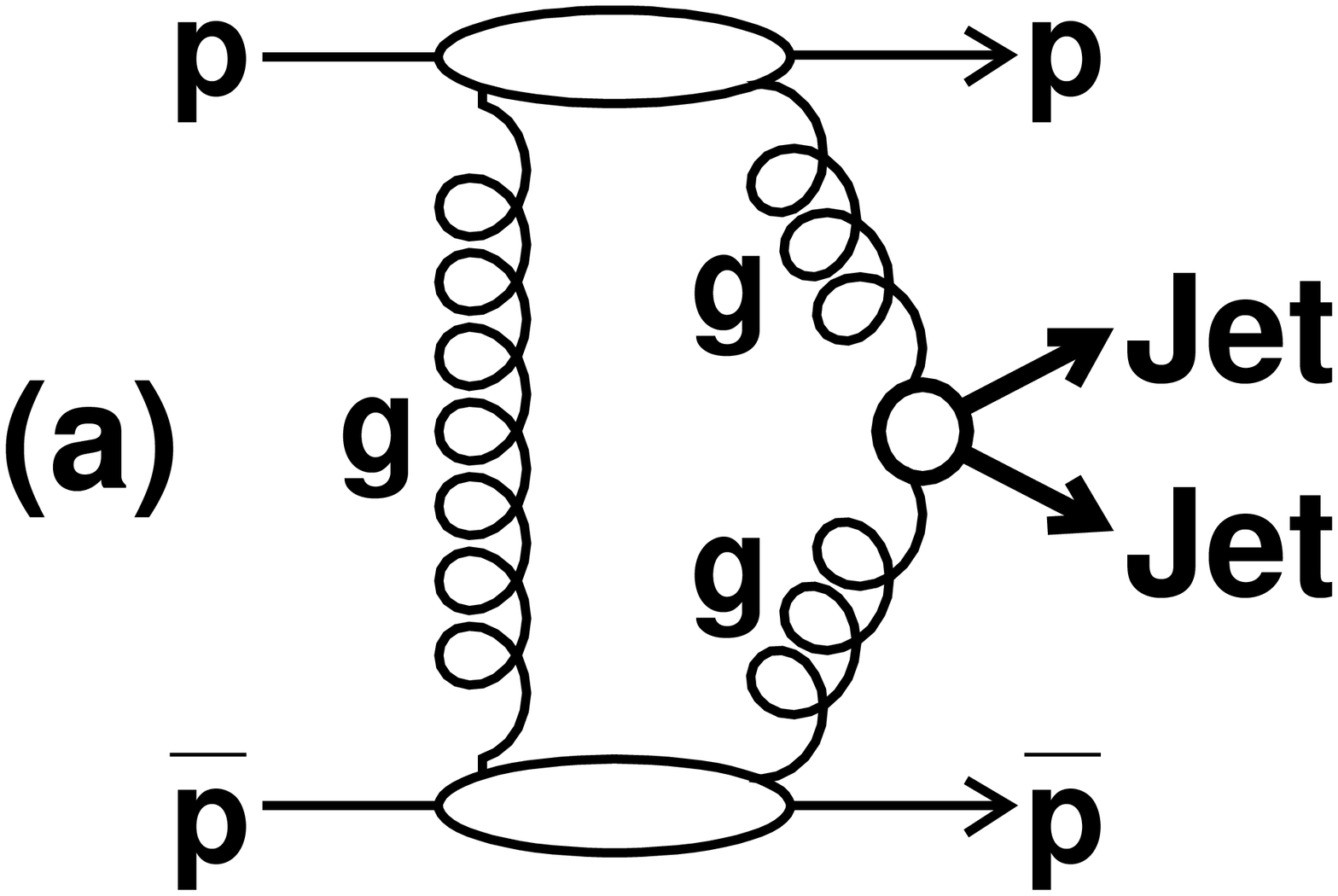}

\includegraphics[width=4cm]{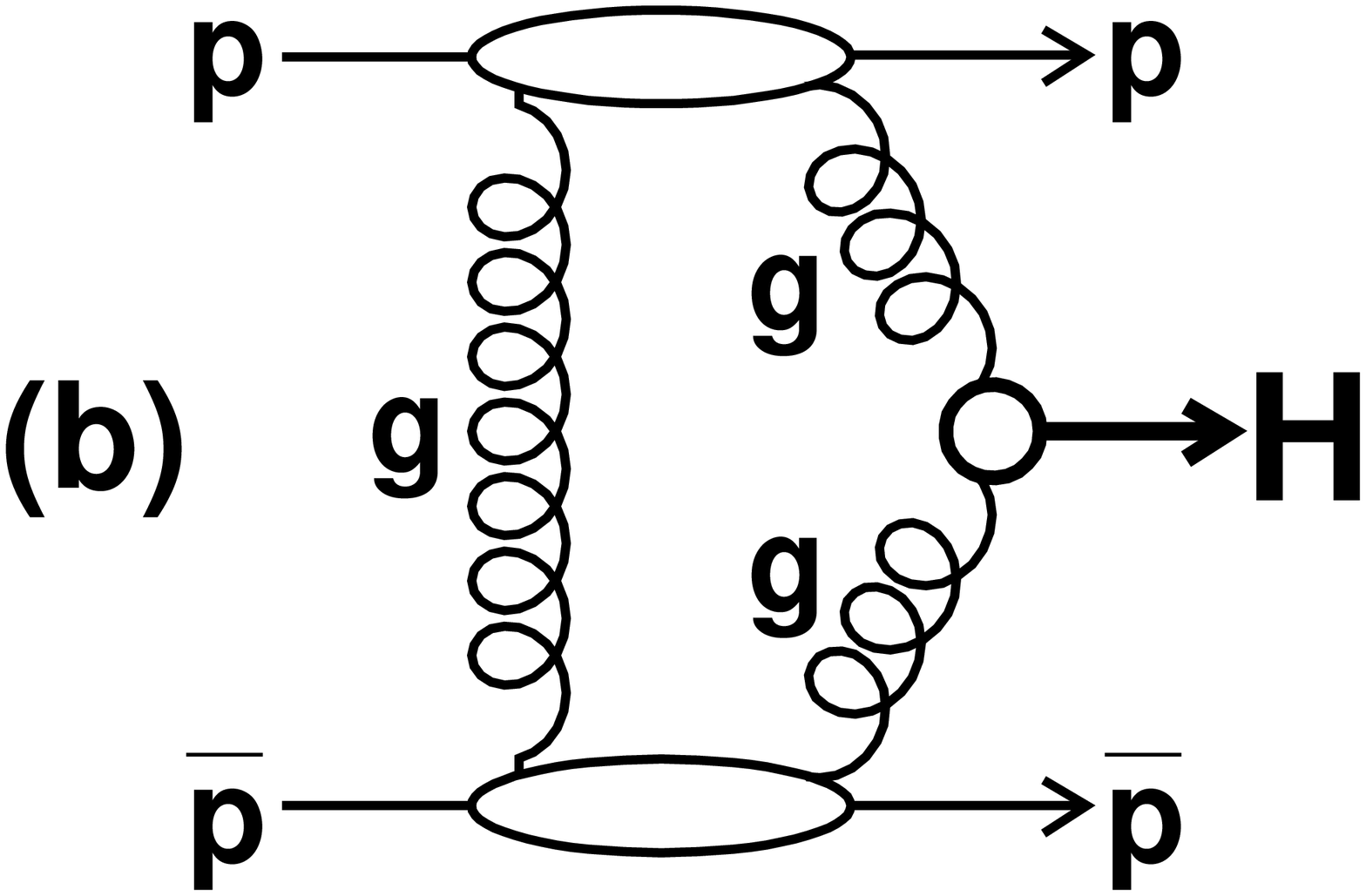}

\vspace*{-16em}
\hspace*{15em}\includegraphics[width=9cm]{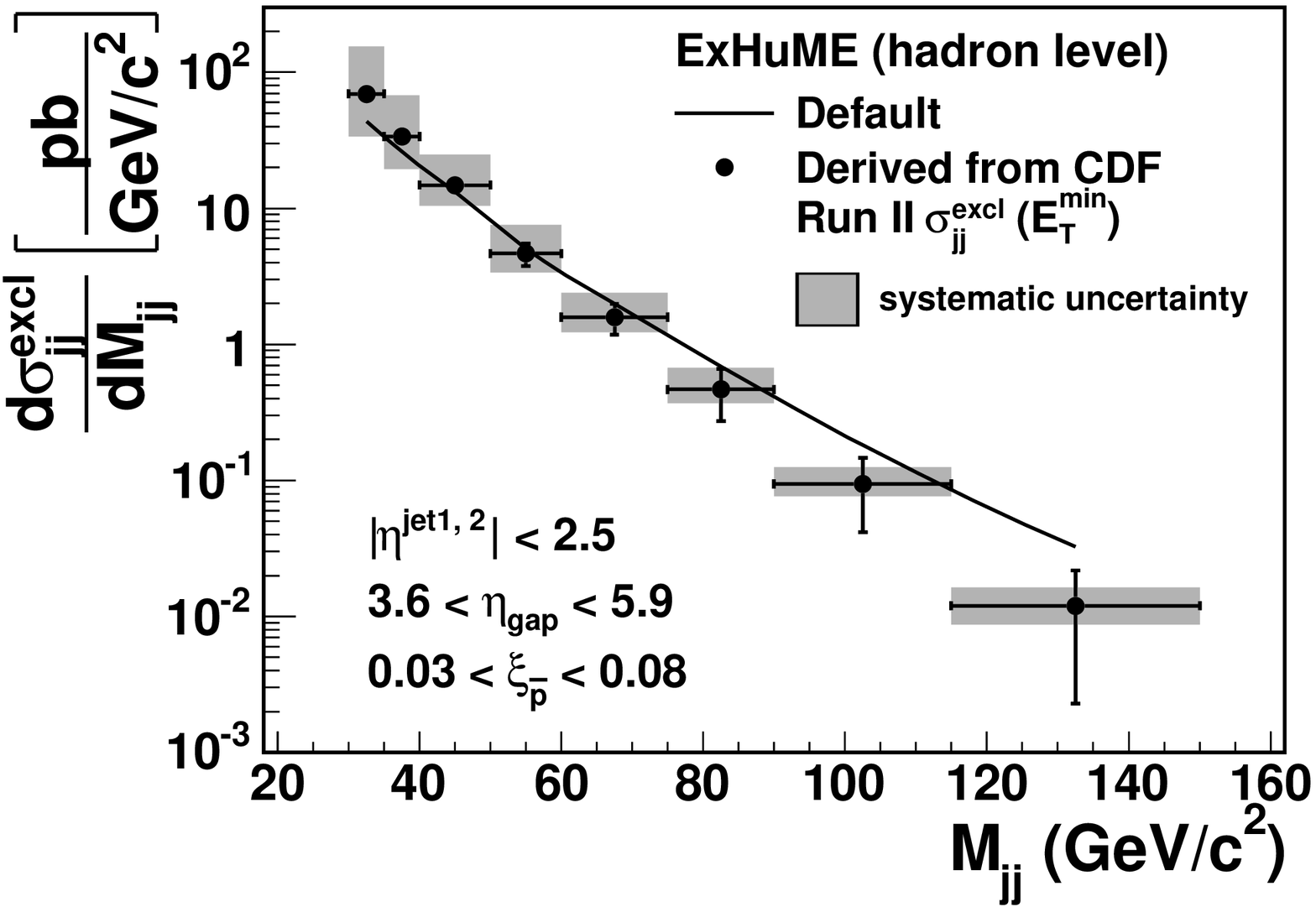}
 \caption{Diagrams for exclusive di-jet (a) and Higgs (b) production, and the {\sc ExHuME}~\cite{exhume} exclusive di-jet differential cross section at the hadron level vs. di-jet mass $M_{jj}$ normalized to measured $\sigma_{jj}^{excl}$ values. The solid curve is the cross section predicted by {\sc ExHuME}.
\label{Fig:xsec_vs_mjj}}
\end{figure}

Central exclusive processes are of the type shown in Fig.~\ref{Fig:xsec_vs_mjj}.
The $p$, $\bar{p}$ emerge intact, with only a small momentum loss, while the
interaction products are emitted into the central region.
Processes of this type may provide a discovery channel for the Higgs boson at the LHC,
as predicted by the KMR model \cite{CEP-KMR}.
Experimental evidence for the validity of this model can be gained from
Tevatron data on CEP of di-jets and di-photons.

The CDF experiment presented final results for the CEP of di-jets \cite{goulianos},
see Fig.~\ref{Fig:xsec_vs_mjj}.
The data favor the KMR model, which is implemented in the Monte Carlo generator
ExHuMe \cite{exhume}.
CDF also reported on evidence for the first observation at hadron colliders of CEP of
di-photons \cite{pinfold}. Three candidate events were observed in $532 ~ \rm{pb}^{-1}$ of CDF run-II data,
in agreement with ExHuMe predictions. Currently it cannot be excluded that some of the
candidate events are $\pi^\circ \pi^\circ$ or $\eta \eta$ events, and further analysis is
on-going.

\subsection{Exclusive dilepton and vector meson production}

Production of exclusive dileptons, of the type
$ \bar{p} p \rightarrow \bar{p} l^+ l^- p$, can occur in two-photon interactions,
$\gamma \gamma \rightarrow l^+ l^-$, and in vector meson photoproduction, $\gamma p \rightarrow
Vp \rightarrow l^+ l^- p$. The former process is to good approximation a pure QED process, with
a cross section known to high precision. At the LHC it can be used for absolute
luminosity determination. The latter process gives access to the gluon content of the proton
and has been studied at HERA for different types of vector mesons.

CDF reported previously the first observation of $\gamma \gamma \rightarrow
e^+ e^-$ production at a hadron collider \cite{pinfold}.
Sixteen events were found in $532 ~ \rm{pb}^{-1}$ of CDF run-II data, where
the electrons have a minimum $E_T$ of 5~GeV and the exclusivity condition requires the absence
of any particle signatures other than the electrons in the full $\eta$ range covered by the
CDF detector, i.e. down to $| \eta| =7.5$. Background pollution was estimated to be
$1.9 \pm 0.3$ events, dominated by events where the $p$ or $\bar{p}$
dissociates. A cross section of $1.6^{+0.5}_{-0.3}(stat) \pm 0.3(syst.)$ pb was obtained, in
good agreement with the prediction from the LPAIR Monte Carlo \cite{lpair}.
The search for exclusive di-muon production in CDF run-II data is on-going. CDF reported
the observation, for the first time at a hadron collider, of $J/\psi$ and $\psi^\prime$
photoproduction candidates \cite{pinfold}.
The background from $\chi_{c} \rightarrow J/\psi + \gamma$(soft),
where the soft photon is not detected, is under study. Search for $\Upsilon$ photoproduction is
likewise under way.

Feasibility studies for exclusive dilepton and vector meson photoproduction were presented by
CMS \cite{Ovyn}. Assuming the cross section predictions of the LPAIR and
STARLIGHT \cite{Nystrand}
Monte Carlo generators,
$\mathcal{O}(700)$ events are expected
for an effective integrated luminosity of single interactions of $100$ pb$^{-1}$.
The largest source of background ($\mathcal{O}(200)$) comes from $p$ dissociative processes.
The number of non-resonant dimuon events will be sufficient to calibrate the
integrated
luminosity to a precision of $\sim 4\%$.
A significant signal for photoproduction of the first three
$\Upsilon$ resonances will be visible.
The mean value of the effective $\gamma p$ center of
mass energy in these events is $\langle W \rangle \sim 2400$~GeV and hence an order of
magnitude larger than for the measurements available from HERA.
The cross section is sensitive to the generalised parton distribution function (GPD) for the
gluons.
The distribution of the four-momentum
squared at the proton vertex, $t$, is sensitive to the two-dimensional distribution of the
gluons in the transverse plane, and has never been measured before for the $\Upsilon$.
Although $t$ cannot be measured directly with CMS,
the measured $p_{T}^{2}$ distribution of the
$\Upsilon$ can be used as effective estimator of the true $t$ distribution.

Theoretical predictions in the color dipole model for the cross section of exclusive
$\Upsilon$ photoproduction at Tevatron and LHC and for the $\Upsilon$ rapidity distribution
were presented in \cite{Watt}. An enduring open question in our understanding of color neutral
gluonic exchanges in perturbative QCD is the absence of experimental evidence for the Odderon,
the $C$-odd partner of the Pomeron. In \cite{Motyka},
the sensitivity at Tevatron and LHC of the
normalized differential cross-section $\,(d\sigma/dp_T^2)/\sigma\,$ for the exclusive
photoproduction of $J/ \psi$ and $\Upsilon$ was discussed.

\subsection{Central exclusive Higgs production}

Interest in diffractive processes at the LHC was boosted in recent years by the realization that
double-Pomeron exchange processes may in fact augment the Higgs discovery reach at the LHC
or render possible a precise measurement of the mass and quantum numbers of the Higgs boson
should it be discovered by traditional searches.
This is the CEP process~\cite{CEP-KMR},
$pp \rightarrow p + \phi + p$, where the plus sign denotes the absence of hadronic
activity between the outgoing protons, which survive the interaction intact, and the
state $\phi$. The final state consists solely of the
scattered protons, which may be detected in forward proton taggers, and the decay
products of $\phi$ which can be detected in the central CMS detector.
Selection rules force the produced state $\phi$ to have $J^{CP} = n^{++}$ with $n =0, 2, ..$.
This process offers hence an experimentally very clean
laboratory for the discovery of any particle with these quantum numbers that couples
strongly to gluons. Additional advantages are the possibility to determine the mass of the
 state
$\phi$ with excellent resolution from the scattered protons alone, independent of its
decay products, and the possibility, unique at the LHC, to determine the quantum numbers of
$\phi$ directly from the azimuthal asymmetry between the scattered protons.

A detailed mapping of the discovery potential of CEP in the MSSM was presented in \cite{Tasevsky}.
In particular CEP of the neutral $CP$-even Higgs bosons $h$ and $H$ and their decays into bottom
quarks has the potential to probe interesting regions of the MSSM parameter space and give access
to the bottom Yukawa couplings of the Higgs boson up to masses of $M_H \le 250$~GeV.
CEP may also give access to the NMSSM Higgs bosons, which would be unique at the LHC
\cite{Forshaw}. Early measurements at the LHC to check the validity of the CEP models for Higgs
production
were discussed in \cite{Khoze}. Forward proton tagging would also provide the possibility for
precision studies of $\gamma p$ and $\gamma \gamma$ interactions at center-of-mass energies
never reached before \cite{VanDerDonckt}.

\subsection{Status of forward physics projects at the LHC}

Both ATLAS \cite{Pilkington} and CMS \cite{VanMechelen} presented the status of their detectors
that extend beyond the central detector volume ($ | \eta | < 5$). Both experiments have
Zero Degree Calorimeters with acceptance for neutral particles for $| \eta | > 8.3$.
Apart from those, their forward instrumentation is quite complementary.
ATLAS is equipped with a dedicated luminosity system consisting of
ALFA, Roman-pot housed tracking detectors at 240~m from the interaction point (IP), which will
peform an absolute luminosity measurement in runs with special LHC optics, and of LUCID,
Cherenkov detectors $(5.6 < | \eta | < 6.0)$ for the primary purpose of luminosity monitoring
during routine LHC data
taking. At the CMS IP, the task of an absolute luminosity determination will be carried out by
an independent experiment, TOTEM, with Roman-pot housed silicon detectors at 220~m distance from
the IP and two tracking telescopes inside of the CMS volume. CMS in addition has the CASTOR
calorimeter which extends the CMS calorimetric coverage to rapidity values of 6.5. CASTOR gives
access to the QCD parton evolution dynamics at very low $x$.
HERA has explored low-$x$ dynamics down to values of a few $10^{-5}$.
At the LHC the minimum accessible $x$ decreases by a factor $\sim 10$ for each
2 units of rapidity. A process with a hard scale of $Q \sim 10$~GeV and within the
acceptance of CASTOR ($\eta \sim 6$) can occur at $x$ values as low as
$10^{-6}$.

In order to give access to CEP Higgs production proton taggers at distances larger than
220/240~m are necessary. The FP420 R\&D collaboration over the last years has investigated the
feasibility of instrumenting the region at 420~m from the ATLAS or CMS IP, within the cryogenic
region of the LHC \cite{Cox}.
The FP420 R\&D report was recently published \cite{FP420report}. It proposes a cryostat re-design to house
the proton taggers which would employ 3-D silicon, an
extremely radiation hard novel silicon technology. Additional
fast timing Cherenkov detectors would be capable of determining, within a resolution of a
few millimeters, whether the tagged proton came from the same vertex as the hard scatter visible
in the central detector. The FP420 proposal is currently under scrutiny in both ATLAS and CMS. If
approved, installation could proceed in 2010, after the LHC start-up.

\subsection{Summary}

Hadron collider experiments have provided a wealth of data on diffraction and vector meson
production in the past, and will continue to do so in the future with the imminent start-up
of the LHC.

\section{Theoretical developments}

QCD collinear factorization method based on the Operator Product
Expansion represents a basis of our understanding of both inclusive
and exclusive hard diffraction processes. It allows to separate
small distance partonic interactions, calculable in the perturbative
QCD, from a large distance dynamics, parameterized by the relevant
parton densities which evolution is described by the DGLAP equation.
This scheme is directly applicable to $ep-$ collisions where the all
order factorization theorems are known for the inclusive diffraction
and for the DVCS and exclusive vector meson production processes. In
the diffraction or low-$x$ limit, where a process energy is much
larger than a hard scale, the description within the leading twist
collinear formalism may be not adequate due to large higher twist
effects and large perturbative corrections. These issues are
addressed in a certain approximation by a dipole model and a
$k_\perp-$ factorization method, which foundations are closely
related to the Balitsky-Fadin-Kuraev-Lipatov (BFKL) formalism for a
resummation of large logarithms of energy. These approaches look
particulary attractive since they allow to incorporate unitarity
corrections in a simple, though still model dependent, way.

K. Kumeri\v cki reported new leading twist QCD analysis of DVCS
\cite{Kumer}. This process considered to be the cleanest way to
access generalized parton densities (GPDs). The level of
sophistication in the description of exclusive DVCS process is
reaching now the one we have for the inclusive DIS. The analysis was
performed including in some approximation the perturbative
corrections up to the next-to-next-to-leading order. This work
relies on an expansion in series of the conformal moments and a
subsequent analytical continuation in the moment index. This
procedure has much in common with the known complex angular momentum
techniques and allows not only to build effective numerical
algorithms but also to get a new insight into the modeling of GPDs.
Good fit to the low-$x$ ZEUS and H1 data was reached. The main
conclusion is that the perturbative corrections seem to be under
control in the HERA kinematic region. Though the convergency of the
perturbative series for DVCS depends much on the input for the gluon
GPD at the initial scale of the QCD evolution. It becomes in general
progressively worse when one approaches the region of very low-$x$.

Exclusive VM production processes provide another exiting insight
into GPD physics. Contrary to the DVCS case the gluon GPD enters in
the description of neutral VM production at the leading order. This
allows to constrain gluon. The quality of data for exclusive VM is
impressive. However, from the theory site it is harder to analyse
them, as both the known NLO perturbative corrections \cite{BM-I-D}
and the existing model predictions for higher twist effects  are
much larger than for DVCS. P. Kroll presented  a comprehensive study
of VM production \cite{Kroll}. In this approach the $t-$ channel
exchange is treated collinearly, in the frame of leading order GPD
formalism, whereas the transition of a virtual photon into VM is
described in a modified factorization scheme, similar to the one
proposed by Lee and Sterman for more simple exclusive reactions
\cite{Li-Sterman}. It allows to describe electroproduction of
transversely polarized VM, which is of higher twist, and it provides
a model for power suppressed contributions in VM production. The
results of \cite{Kroll} reproduce fairly well many features of the
data from both the fixed target and the HERA collider experiments.

M. Diehl presented an overview of some recent developments (mostly
not presented at the workshop) in the theory and phenomenology of
GPDs. It concerns both the GPD modeling and the extraction of GPDs
from the exclusive reactions data \cite{Diehl}.

It is known \cite{Bartels:1998ea} that the twist 4 contribution to
the longitudinal part of inclusive diffraction structure function
$F_L^D$ dominates the leading twist 2 one at $\beta\to 1$ (i.e. at
small diffractive masses, $M^2\ll Q^2$); the twist 2 vanishes as
$\beta\to 1$, while its the higher twist counterpart not. K.
Colec-Biernat reported the analysis \cite{Biernat} of the
diffractive HERA data which includes this important twist 4
contribution to $F_L^D$ within the model provided by
\cite{Bartels:1998ea}. The most important result of this study in
comparison to the standard QCD leading twist fit is that the higher
twist significantly enhances $F_L^D$ at $\beta>0.7$ and leads to the
extracted twist 2 gluon diffractive distribution which is strongly
peaked near $\beta=1$. It calls for a direct measurement of $F_L^D$
at large $\beta$ in HERA experiments.

Hard diffraction in hadron collisions in general and especially a
description of central exclusive production processes represents the
real challenge for a theory. There is no factorization theorem for
these reactions. Therefore we relay here more on a physical
intuition and on a phenomenological experience rather than on some
systematic expansion. Valery Khoze presented an approach developed
by the Durham group \cite{Khoze}. It was used for the predictions of
CEP of Higgs boson; both in the Standard Model \cite{CEP-KMR} and
beyond \cite{Tasevsky,Forshaw}. The cross section of CEP di-jet
production predicted within this approach agrees with recent CDF
measurement \cite{goulianos}. Valery stressed that the method
involves a number of nontrivial steps which may be tested to some
extend separately, studying various diffraction processes in early
LHC data. In particular it involves an assumption that the
rescattering effects are related with the soft pomeron exchanges of
the eikonal type and can be parameterized using elastic and
quasy-elastic proton-proton scattering data. However, more
complicated rescattering mechanisms that involve the pomeron loops
were found in \cite{Bartels:2006ea}. J. Miller presented the model
calculation of such enchanced pomeron loop diagrams. Effectively it
results in a much larger prediction for the gap survival suppression
\cite{miller}. More discussion of these issues were given by M.
Diehl \cite{Diehl1}.

J. Nystrand discussed the exclusive VM production in
ultra-peripheral photon collisions at the LHC and Tevatron
\cite{Nystrand}, including an interesting interference effect
between the photon-pomeron and the pomeron-photon production
mechanisms. The outlook for the LHC is promising. The theory
predictions depend strongly on the input for VM photoproduction
amplitude in the energy region not explored so far in HERA
experiments. These processes should be a good testing ground for
both our knowledge of the gluon at low-x and for our understanding
of the production mechanism.

Another exiting possibility is related with the photon-photon
induced processes at the LHC. O. Kepka showed that the potential of
$\gamma\gamma$ collisions can be used to put stronger constrains on
the anomalous $WW\gamma$ triple gauge boson vertex \cite{Kepka}.

G. Watt discussed the exclusive photo- and electroproduction
processes within the color dipole picture \cite{Watt}. In this case
it requires a modeling of the impact parameter dependent
dipole-proton cross section. Good description of both the DIS and
exclusive VM production HERA data was achieved. It allows to
estimate the impact parameter dependent saturation scale parameter,
which found to be less than $0.5\, \mbox{GeV}^2$ in the HERA
kinematic regime. Using these results the predictions for exclusive
VM and $Z^0$ boson production processes at LHC were given. K. Tuchin
presented arguments that saturation effects may be very important
for the hard diffractive processes in pA collisions \cite{Tuchin}.
In particular, diffractive hard gluon production in onium- nucleus
scattering was studied.

The other line of development is related with the $t-$ channel
picture for diffractive processes in perturbative QCD. Its main
building block is the Reggeized gluon exchange. Two interacting
gluons in the color singlet state form the hard pomeron. It is
described by the BFKL equation and is known at present with the
next-to-leading $\log(1/x)$ accuracy (NLA). The impact factors are
the other important ingredients which represent the coupling of the
Reggeized gluons to the external particles. It seems to be extremely
important to understand the role of subleading energy logarithms in
the BFKL approach. A. Papa \cite{Papa} considered the forward
amplitude for the production of two VM in the collision of two
virtual photon, $\gamma^*\gamma^*\to VV$. This amplitude can be
presented in a closed analytical form since the $\gamma^*\to V$
impact factor is known at the NLA accuracy. The sensitivity of the
amplitude to the scheme choice, the renormalization and the energy
scales was systematically analyzed. Stable results for the amplitude
were found applying optimization of the perturbative expansion. The
optimal value of the scales turns out to be much larger than the
kinematical one, given by $\gamma^*$ virtuality. This should be a
manifestation of large corrections subleading to NLA. These effects
were studied approximately, applying known collinear improvement
method to the kernel of BFKL equation. It is interesting that after
collinear improvement the optimal value of the renormalization
scales gets much closer to the kinematical one. This may indicate
that collinear improvement method catches the essential sub-NLA
dynamics.

A excellent probe of the BFKL dynamics may be given by
Mueller-Navelet jets at hadron colliders. A. Sabio-Vera presented
the analysis for the azimuthal angle decorrelation of the jets and
provided estimates for the Tevatron and LHC \cite{Vera}.

L. Motyka addressed the problem of unitarity corrections for the
high energy baryon scattering in perturbative QCD \cite{Motyka1}.
The consideration is relied on the model for a light cone wave
function that encodes the nontrivial baryon quantum numbers. The
impact factors describing the coupling of a baryon to the states of
up to four Reggeized gluons and their energy evolution are studied.
New structures in the interaction of a baryon with Reggeized gluons
were found. It may indicate that the pattern of unitarisation for
baryon amplitudes is more complex as compared to scattering of the
color dipole.


\begin{footnotesize}

\end{footnotesize}


\begin{thebibliography}{99}
\bibitem{url} Slides: \\
\verb$http://indico.cern.ch/contributionDisplay.py?contribId=8&sessionId=8&confId=24657$

\bibitem{regge}
P.D.B.~Collins, {\em An Introduction to {Regge}
Theory and High Energy Physics,\rm{ Cambridge University Press, Cambridge}}, 1977.

\bibitem{dl}
A. Donnachie, P.V. Landshoff, Phys. Lett. {\bf B296} 227 (1992);\\
J. Cudell {\it et~al.},   Phys. Rev. {\bf D61} 034019 (2000); Erratum,  Phys. Rev. {\bf D63} 059901 (2001).

\bibitem{arneodo} M. Arneodo, M. Diehl, {\it Diffraction for non-believers}, hep-ph/0511047 (2005)

\bibitem{fact}
L. Trentadue and G. Veneziano, Phys. Lett. {\bf B323} 201 (1994);\\
J.C. Collins, Phys. Rev. {\bf D57} 03051 (1998); Erratum, ibid.  {\bf D61}
019902 (2000);\\
A. Barrera and D.E.Soper, Phys. Rev. {\bf D53} 06162 (1996).

\bibitem{LRG}
H1 Coll., A. Aktas {\it et~al.}, Eur. Phys. J. {\bf C48} 715 (2006);\\
H. Abramowicz, Int. J. Mod. Phys. A {\bf S1b}, 495 (2000).

\bibitem{ptagging}
H1 Coll., A. Aktas {\it et~al.}, Eur. Phys. J. {\bf C48} 749 (2006);\\
ZEUS Coll., S. Chekanov {\it et~al.}, Eur. Phys. J. {\bf C38} 43 (2004);\\
ZEUS Coll., S. Chekanov {\it et~al.}, Eur. Phys. J. {\bf C25} 169 (2002);\\
ZEUS Coll., J. Breitweg {\it et~al.}, Eur. Phys. J. {\bf C1} 81 (1997).

\bibitem{mx}
\verb$http://www-h1.desy.de/h1/www/publications/H1_sci_results.shtml$, H1-prelim-06-014;\\
ZEUS Coll., S. Chekanov {\it et~al.}, Eur. Phys. J. {\bf C25} 169 (2002).

\bibitem{mx1}
ZEUS Coll., S. Chekanov {\it et~al.}, Nucl. Phys. {\bf B713}, 3 (2005).

\bibitem{mx2}
ZEUS Coll., S. Chekanov {\it et~al.}, DESY-08-011, submitted to Nucl. Phys. {\bf B}.

\bibitem{marta}
\verb$http://www-zeus.desy.de/public_results/groupsearch.php$,  ZEUS-pub-08-010.

\bibitem{ingsc}
G. Ingelman and P.E. Schlein, Phys. Lett. {\bf B152} 256 (1985).

\bibitem{marcella}
\verb$http://www-zeus.desy.de/public_results/groupsearch.php$,  ZEUS-pub-08-013.

\bibitem{h1lrg}
H1 Coll., A. Aktas {\it et~al.}, Eur. Phys. J. {\bf C48} 715 (2006).

\bibitem{h1fps}
H1 Coll., A. Aktas {\it et~al.},  Eur. Phys. J. {\bf C48} 749 (2006).

\bibitem{h1pdf}
H1 Coll., A.Aktas {\it et~al.}, JHEP {\bf 0710} 042 (2007).

\bibitem{h1dvcs}
H1 Coll., A. Aktas {\it et~al.},  Phys. Lett. {\bf B659} 796 (2008).

\bibitem{ruspa}
M. Ruspa, these proceedings.
\bibitem{newmann}
P. Newman, these proceedings.
\bibitem{laycock}
P. Laycock, these proceedings.
\bibitem{favart}
L. Favart, these proceedings.
\bibitem{ukleja}
Justyna Ukleja, these proceedings.
\bibitem{malka}
J. T. Malka, these proceedings.
\bibitem{slominski}
W. Slominski, these proceedings.
\bibitem{cerny}
K. Cerny, these proceedings.
\bibitem{janssen}
X. Janssen, these proceedings.
\bibitem{levy}
A. Levy, these proceedings.
\bibitem{rinaldi}
L. Rinaldi, these proceedings.


\bibitem{Bj} J.D. Bjorken, Phys. Rev. D {\bf 47} (1993) 10111;
A.B. Kaidalov et al., Eur. Phys. J. C {\bf 21} (2001) 521
\bibitem{CF}
V.A. Khoze, A.D. Martin, M.G. Ryskin, Eur. Phys. J. C {\bf 55} (2008) 363
\bibitem{Ssquared}
V.A. Khoze, A.D. MArtin, M.G. Ryskin, Phys. Lett. B {\bf 643} (2006) 93;
E. Gotsman, E. Levin, U. Maor, E. Naftali, A. Prygarin, hep-ph/0511060
\bibitem{miller} J.S. Miller, Eur. Phys. J. C {\bf 56} (2008) 39
\bibitem{CEP-KMR} V.~A.~Khoze, A.~D.~Martin, M.~G.~Ryskin., Eur.Phys.J C19 (2001) 477, erratum C20
(2001) 599
\bibitem{goulianos} K. Goulianos, these proceedings
\bibitem{vilela} A. Vilela Pereira, these proceedings
\bibitem{exhume} J. Monk, A. Pilkington, Comput. Phys. Commun. {\bf 175} (2006) 232
\bibitem{pinfold} J. Pinfold, these proceedings
\bibitem{lpair} J.A.M. Vermaseren, Nucl. Phys. B {\bf 229} (1983) 347
\bibitem{Nystrand} J. Nystrand, these proceedings
\bibitem{Ovyn} S. Ovyn, these proceedings
\bibitem{Watt} G. Watt, these proceedings
\bibitem{Motyka} L. Motyka, these proceedings
\bibitem{Tasevsky} M. Tasevsky, these proceedings
\bibitem{Forshaw} J. Forshaw, these proceedings
\bibitem{Khoze} V. Khoze, these proceedings
\bibitem{VanDerDonckt} M. Van Der Donckt, these proceedings
\bibitem{Cox} B. Cox, these proceedings
\bibitem{VanMechelen} P. Van Mechelen, these proceedings
\bibitem{Pilkington} A. Pilkington, these proceedings
\bibitem{Kepka} O. Kepka, these proceedings
\bibitem{FP420report} FP420 collab, M.G. Albrow et al.,
{\it The FP420 R\&D Project: Higgs and New Physics with forward protons at the LHC},
arXiv:0806.0302 [hep-ex] (2008)

\bibitem{Kumer} K. Kumeri\v cki, these proceedings

\bibitem{BM-I-D} A.V.~Belitsky and D.~Mueller,
  Phys.\ Lett.\  B {\bf 513} (2001) 349;\\
  D.Yu.~Ivanov, L.~Szymanowski and G.~Krasnikov,
  JETP Lett.\  {\bf 80} (2004) 226;\\
   D.Yu.~Ivanov, A.~Schafer, L.~Szymanowski and G.~Krasnikov,
  Eur.\ Phys.\ J.\  C {\bf 34} (2004) 297;\\
  M.~Diehl and W.~Kugler,
  Eur.\ Phys.\ J.\  C {\bf 52} (2007) 933

\bibitem{Kroll} P. Kroll, these proceedings


\bibitem{Li-Sterman}
  H.n.~Li and G.~Sterman,
  Nucl.\ Phys.\  B {\bf 381} (1992) 129


\bibitem{Diehl} M. Diehl, Some News about Generalized Parton Distributions,
these proceedings

\bibitem{Bartels:1998ea}
  J.~Bartels, J.R.~Ellis, H.~Kowalski and M.~Wusthoff,
  Eur.\ Phys.\ J.\  C {\bf 7} (1999) 443

\bibitem{Biernat} K. Golec-Biernat, these proceedings

\bibitem{Bartels:2006ea}
  J.~Bartels, S.~Bondarenko, K.~Kutak and L.~Motyka,
  Phys.\ Rev.\  D {\bf 73} (2006) 093004


\bibitem{Diehl1} M. Diehl, Diffraction and Forward Physics: From HERA to LHC,
these proceedings

\bibitem{Tuchin} K. Tuchin, these proceedings



\bibitem{Papa} A. Papa, these proceedings

\bibitem{Vera} A. Sabio-Vera, these proceedings

\bibitem{Motyka1} L. Motyka, Proton elastic impact factors for two, three, and four gluons; these proceedings


\end{thebibliography}
\end{document}